# A nanogapped hysteresis-free field-effect transistor

Jiachen Tang, Luhao Liu, Yinjiang Shao, Xinran Wang, Yi Shi and Songlin Li[a]

*National Laboratory of Solid-State Microstructures, Collaborative Innovation Center of Advanced Microstructures, and School of Electronic Science and Engineering, Nanjing University, Nanjing, Jiangsu 210023, China*

[a] *Author to whom correspondence should be addressed:* sli@nju.edu.cn

**Abstract:** We propose a semi-suspended device structure and construct nanogapped, hysteresis-free field-effect transistors (FETs), based on the van der Waals stacking technique. The structure, which features a semi-suspended channel above a submicron-long wedge-like nanogap, is fulfilled by transferring ultraclean BN-supported $MoS_2$ channels directly onto dielectric-spaced vertical source/drain stacks. Electronic characterization and analyses reveal a high overall device quality, including ultraclean channel interfaces, negligible electrical scanning hysteresis, and Ohmic contacts in the structures. The unique hollow FET structure holds the potential for exploiting reliable electronics, as well as nanofluid and pressure sensors.

Two-dimensional (2D) materials have attracted significant interest from the scientific community due to their unique physical dimensionality and prospects for various technological applications, such as chemical and biological nanofluidic and pressure sensors,[1-8] as well as post-silicon electronics.[9-13] In practice, however, electronic devices consisting of 2D channels are subject to issues in stability and reliability due to scanning hysteresis, arising from the ubiquitous traps in supporting substrates.[14,15] Physical suspension and clean dielectric encapsulation (e.g., with hexagonal boron nitride, BN)[16-19] are proven to be effective strategies to eliminate the adverse electronic hysteresis in devices, but the fully suspended or encapsulated device structures normally suffer from shortcomings in mechanical strength or isolation of the functional sensing areas. Hence, a device structure compatible with both features of openness and interface passivation is highly sought after.

In this letter, we address the above issue by proposing a nanogapped, hollow field-effect transistor (FET) structure with one-side-open, one-side-encapsulated Janus channels, based on the van der Waals stacking technique. The structure is realized by assembling the ultraclean BN-supported $MoS_2$ channel directly onto the BN-spaced vertical source/drain stack and features a semi-suspended configuration with a submicron-long, wedge-shaped nanogap between the channel and substrate. Electronic characterization reveals a high overall device quality, including negligible electronic scanning hysteresis, Ohmic contacts, and band-like conduction behavior. Besides generally high current switching ratios over $10^5$, we also achieve high ON-state currents over 30 μA/μm in sub-micron $MoS_2$ FETs. Furthermore, this device structure is also applicable for alternative channels with ambipolar conduction and for cascading between multiple devices. A prototype voltage inverter, with a static voltage gain of 9, is also demonstrated.

The main preparation steps of the nanogapped FETs are shown in Fig. 1(a). At first, a trilayer Au/BN/Au stack with $CF_4$ etched sidewall is defined by multiple steps including electron beam lithography (EBL), thermal evaporation, and self-aligned BN etching. Here, the BN middle layer is used as the spacer to isolate the top and bottom Au electrodes and, thus, its thickness is responsible for device leakage. We note that for simplicity, the 2 nm Ni adhesive layers below the Au electrodes are omitted in the diagrams. Then, a BN-supported $MoS_2$ channel is transferred onto the sidewall with each end contacting one of the two stacked Au electrodes, in order to form the nano-wedge gapped structure. Thus, the height of the nano-wedge is the sum of the thicknesses of the BN spacer and top Au electrode, which can be finely controlled down to 10 nm, making it particularly suitable for nanofluidic trace detection. Finally, a top gate is defined atop the BN supporter (i.e., gate dielectric), to complete the entire device. The detailed preparation steps and device images can be found in supplementary material, Fig. S1. Figure 1(b) shows the top-view optical image for a typical nanogapped hollow FET. For clarity, the $MoS_2$ channel, BN gate dielectric, and sidewall are outlined by dotted white, blue, and red lines, respectively. The realistic functional region is just located on the right to the sidewall.

To shed light on the formation mechanism and realistic configuration inside the hollow structure, its cross-section was examined by transmission electron microscopy (TEM) [Fig.

1(c)]. Two features can be seen. First, the device formation relies on the gentle mechanical bending of the top gate stacks around the two contacting points and the slight deformation at the end of the drain electrode, resulting from the mechanical force applied during the van der Waals stacking. All the physical deformation and adaption of each layer contribute to the structural steadiness of the devices and the high contact quality between the channel and electrodes, as will be discussed below. Apart from the contacting parts, the realistic length of the suspended $MoS_2$ channel is about 400 nm [Fig. 1(c)]. Second, the top surface of the channel is attached firmly to the BN dielectric while the bottom surface is suspended, leading to the one-side-open, one-side-encapsulated channel geometry. Such a Janus-faced geometry can not only preserve the openness required by nanofluid sensing but also avoid contamination by substrate traps, holding potential for versatile applications in reliable electronics. A schematic 3D structure and the elemental mapping for the channel region are further given in Figs. 1(d) and 1(e), respectively, to better understand the nanogapped device structure. Surface morphological characterization indicates that the thicknesses of the BN spacer, $MoS_2$ channel, and BN gate dielectric are 10, 4, and 15 nm, respectively. The channel corresponds to a 6-layer $MoS_2$.

The electrical performance of the semi-suspended $MoS_2$ FET was characterized under a dark high-vacuum (<$10^{-3}$ Pa) condition. The corresponding transfer curves are plotted in Fig. 1g. In the scanning range of gate voltage ($V_{gs}$) from -6 to 7 V, the 400-nm short-channel FET exhibits an effective switching property with the current switching ratio ~$10^5$ at drain-source voltage ($V_{ds}$) of 0.1 V, without showing obvious behavior of drain-induced barrier lowering. The ON-state drain-source current ($I_{ds}$) reaches up to 30 µA/µm at $V_{ds}$ = 0.5 V, featuring a high value among the counterparts under similar biasing conditions. The cleanness of this semi-suspended FET is totally analogous to the fully encapsulated devices, as manifested in its negligible scanning hysteresis, implying the minimal effects from the substrate traps, which represents an outstanding merit for exploiting reliable sensing and switching devices. In addition, the gate leakage is less than 0.2 pA, implying the excellent isolation of the BN spacer.

As a reference experiment, we also fabricated the semi-suspended FET with 20 nm $Al_2O_3$, rather than BN, as the gate dielectric, to further elucidate the importance of the surface conditions in dielectric layers. As shown in supplementary material, Fig. S2, significant scanning hysteresis arises in its transfer curves, indicating the existence of a large density of trap states in the atomic layer deposition prepared high-k gate dielectrics due to the presence of unsaturated surface dangling bonds therein. Nevertheless, the current switching ratios of the device are enhanced to $10^8$, because of the enhanced coupling capacitance from the high-k gate dielectrics. Hence, well passivated high-k dielectrics would be required for the highly shortened channels.

To gain further insight into the device quality of the semi-suspended FET, we performed cryogenic electrical transport characterization. As shown in Fig. 2(a), the device exhibits a band-like transport characteristic because the transfer curves show generally higher currents at lower temperatures ($T$s). At $V_{gs}$ = 7 V, $I_{ds}$ increases from 70 to 180 μA when $T$ is reduced from 300 to 8 K. We note that the negative $I_{ds}$-$T$ correlation is seldom observed in two-terminal FETs due to the presence of high contact resistance ($R_C$) in most 2D channel FETs. This observation suggests the presence of high-quality metal/semiconductor contacts defined by the van der Waals stacking.

The FET exhibits a field-effect induced metal-insulator transition (MIT)[20] around $V_{gs}$ = 2.2 V, at which all transfer curves intersect. Such MIT behavior can be seen more clearly in the plot of conductivity ($\sigma$) versus $T$ at different gating conditions. When $V_{gs}$ < 2.2 V, the channel is electrically insulating (i.e., in the depletion state) and $\sigma$ exhibits positive $T$ dependence. In contrast, when $V_{gs}$ > 2.2 V, the channel becomes metallic (i.e., in the inversion state) and the $\sigma$-$T$ dependence turns negative. Interestingly, at $V_{gs}$ = 2.2V the MoS$_2$ channel enters a quantum critical state with its sheet conductivity ~$e^2$/h.

To evaluate the injection barrier ($\Phi_B$) at contacts, we employed the 2D thermionic emission model that is commonly used for extracting the effective thermionic $\Phi_B$ in FETs.[21-26] As such, the device current can be written as

$$I_{ds}(V_{gs}) = WA^*T^{3/2}\exp(-\frac{q\phi_B}{k_BT})[\exp(\frac{qV_{ds}}{nk_BT})-1], \quad (1)$$

where $W$ is the channel width, $A^*$ is the modified Richardson constant, $q$ is the elementary charge, $k_B$ is the Boltzmann constant, and $n$ is the ideality factor. The inset of Fig. 2(b) shows the Arrhenius plot of ln($I_{ds}/T^{3/2}$) versus 1000/$T$ at $V_{ds}$ = 0.5 V. In such plots, the slopes correspond to the negative values of effective $\Phi_B$ at different $V_{gs}$. They evolve from negative to positive at the transition point of $V_{gs}$ ~ 0.4 V. The extracted $\Phi_B$ values are plotted versus $V_{gs}$ in Fig. 2(b), which reveals a flat-band voltage ($V_{FB}$) near 0.4 V and an effective thermionic $\Phi_B$

~5 meV. The low barrier value corroborates the high quality of the van der Waals contacts.

For sensing applications, atomically thin monolayer (1L) channels are expected to be more sensitive than multilayers. To verify the applicability of the hollow device configuration to atomically thin channels, we also integrated 1L MoS$_2$ as the semi-suspended channel [Fig. 2(c)]. A remarkable point is that the current switching characteristic is greatly enhanced to $10^8$ at $V_{ds}$ = 0.5 V, arising from the suppression of the short-channel effect from the further reduction in channel thickness. A small scanning hysteresis arises in the subthreshold regime for $V_{gs}$ around -1.5 V, possibly resulting from the release of residual surface absorbates due to the enhanced sensitivity to charge variation. The ON-state $I_{ds}$ is 3 µA/µm at $V_{ds}$ = 0.5 V. The excellent electrical characteristics confirm the broad applicability of this device structure.

Apart from the negative-type MoS$_2$ channels, we then verify the feasibility of integrating channels with other conduction polarities. In Fig. 2(d), we present a semi-suspended FET consisting of a bilayer (2L) WSe$_2$ channel, in which the BN spacer and gate dielectric are 25 and 10 nm in thickness. As expected, the device exhibits ambipolar conduction behavior with switching ratios up to $10^8$ upon gating. More importantly, the negligible scanning hysteresis is reproducibly observed in the transfer curves, suggesting the universality of this device construction strategy for different materials, as well as the possibility of further function extension, such as ambipolar electronic devices.[27,28]

Due to the limit of device configuration, it is difficult to directly assess the $R_C$ in the individual semi-suspended FETs. For this reason, we adopted the Y-function method[29,30] to estimate the $R_C$ (supplementary material, Fig. S3), which, at the strong inversion regime, are 1.37 and 1.42 kΩ·µm for 8-layer (8L) and 6-layer (6L) MoS$_2$ channels, respectively, and are comparable to the reported high-quality devices after contact optimization.[31,32] In addition, to check the accuracy of the Y-function method, we also prepared multiple FETs with van der Waals MoS$_2$/Au contacts and extract accurate $R_C$ through the well-accepted transfer-length method (TLM). Figures 3(a)−3(d) show the serial optical images for the channel length varied FETs at different stages. For accuracy, all the channels were etched into a uniform strip with the same width [Fig. 3(c)]. Figure 3(e) shows the total resistance ($R_{total}$) versus channel length at four charge densities ($n_{2D}$). For the trilayer (3L) MoS$_2$/Au van der Waals contacts, $R_C$ is reduced from 3.5 to 2.3 kΩ·µm, as $n_{2D}$ increases from 1.1 to 4.4×10$^{13}$ cm$^{-2}$. The values are slightly higher than those from the Y-function method, likely resulting from the reduced channel thickness.[33] For clarity, the device characteristics for all measured semi-suspended FETs are summarized in Table 1.

Finally, we demonstrate a simple application by fabricating a prototype of voltage inverter with two connected semi-suspended FETs. Figures 4(a) and 4(b) show the schematic circuit diagrams and optical image for the inverter, in which one serves as the pull-up resistor controlled by the bias voltage ($V_{\text{bias}}$) and the other works as the switching unit. After proper wiring and biasing, the inverter works well with its voltage transfer characteristic shown in Fig. 4(c). Under all supply biasing conditions, the output voltage ($V_{\text{out}}$) changes from its high ($V_{\text{DD}}$) to low levels (ground), in response to the variation of input voltage ($V_{\text{in}}$) from low to high voltage levels. Its voltage gain, defined as the absolute value of the differential of $V_{\text{out}}$ with respect to $V_{\text{in}}$, is then estimated in Fig. 4(d). In line with conventional devices, the voltage gain grows with increasing $V_{\text{DD}}$, and reaches a maximum value of ~9 at $V_{\text{DD}}$ = 5 V. The device performance is expected to be further improved if the gate dielectrics can be further optimized with higher capacitive coupling abilities.

In conclusion, we have developed a nano-gapped, semi-suspended FET structure consisting of 2D semiconductor channels. Such a unique device structure can minimize the adverse effect of substrate charge traps, which results in a high overall device cleanness and electronic performance, including high on-state current over 30 µA/µm, switching ratio around $10^8$, and negligible electronic scanning hysteresis. Moreover, the device structure is applicable to materials with different conduction polarities. A basic electronic function as voltage inverter is also demonstrated. Together with the unique nanogapped structure, the outstanding overall device quality render it suitable for exploiting highly reliable chemically and biologically nanofluid sensing and advanced logic applications.

**SUPPLEMENTARY MATERIAL**

See the supplementary material for the detailed device fabrication steps, description of Y-function method, electronic characteristics of semi-suspended FET with $Al_2O_3$ dielectrics, and individual elemental mappings around the channel area.

**ACKNOWLEDGMENTS**

This work was supported by the National Key R&D Program of China (2021YFA1202903), the National Natural Science Foundation of China (61974060 and 61674080), and the Innovation and Entrepreneurship Program of Jiangsu province.

## AUTHOR DECLARATIONS

### Conflict of Interest

The authors declare that they have no conflicts of interests.

## DATA AVAILABILITY

The data that support the findings of this study are available from the corresponding author upon reasonable request.

# Figure captions

**FIG. 1.** (a) Simplified processing flow for the nanogapped, semi-suspended FETs. (b) Top-view optical image for an as-prepared device. (c) Cross-sectional TEM image. (d) Schematic 3D illustration. (e) Elemental mapping around the semi-suspended channel. (f) Surface morphologies taken by AFM to identify the thicknesses of the BN spacer, $MoS_2$ channel and BN gate dielectric. The channel corresponds to 6 layers of $MoS_2$. (g) Top: transfer curves in the linear and logarithmic plots at different $V_{ds}$ from 0.1 to 0.5 V. Bottom: gate leakage recorded at $V_{ds}$ = 0.5V.

**FIG. 2.** (a) Transfer curves at various $T$ values ranging from 8 to 300 K. Inset: σ versus $T$ under different gating conditions. (b) Calculated values for effective thermionic $\Phi_B$ versus $V_{gs}$ at different $V_{ds}$ values from 0.1 to 0.5V. Inset: Arrhenius plots of $\ln(I_{ds}/T^{3/2})$ versus $1000/T$. (c) Transfer characteristics of a 1L $MoS_2$ semi-suspended FET. Inset: optical image of the 1L $MoS_2$ FET with channel width ~2.5 μm. (d) Transfer characteristics for a 2L $WSe_2$ semi-suspended FET. Inset: optical image of the 2L $WSe_2$ FET with channel width ~1 μm.

**FIG. 3.** $R_C$ extraction using the transfer-length method for the van der Waals stacked 3L $MoS_2$/Au contacts, where 15-nm $HfO_2$ is used as top dielectric. (a)−(d) Serial optical images taken at different stages during device fabrication (e) Plot of total resistance versus channel length for $R_C$ extraction.

**FIG. 4.** (a) and (b) show the schematic circuit diagrams and optical image for a voltage inverter comprising two semi-suspended FETs. (c) Voltage transfer characteristics ($V_{out}$ versus $V_{in}$) for the inverter operated at different $V_{DD}$ from 1 to 5 V. (d) Corresponding voltage gain ($dV_{out}/dV_{in}$) to (c).

# Table

**Table 1.** Summary for the electronic characteristics of semi-suspended FETs.

| Channel material | Channel thickness | Channel length | Channel width | Switching ratio | Max. $I_{ds}$ ($V_{ds}$=0.5V) | Min. $R_C$ |
|---|---|---|---|---|---|---|
| $MoS_2$ | 8L | 1 μm | 1.5 μm | $3\times10^5$ | 30 μA/μm | 1.37 kΩ·μm |
| $MoS_2$ | 6L | 0.4 μm | 3 μm | $6\times10^4$ | 30 μA/μm | 1.42 kΩ·μm |
| $MoS_2$ | 3L | 0.24 μm | 0.6 μm | $10^9$ | 60 μA/μm | 2.3 kΩ·μm |
| $MoS_2$ | 1L | - | 2.5 μm | $10^8$ | 3 μA/μm | - |
| $WSe_2$ | 2L | - | 1 μm | $10^8$ | 2 μA/μm | - |

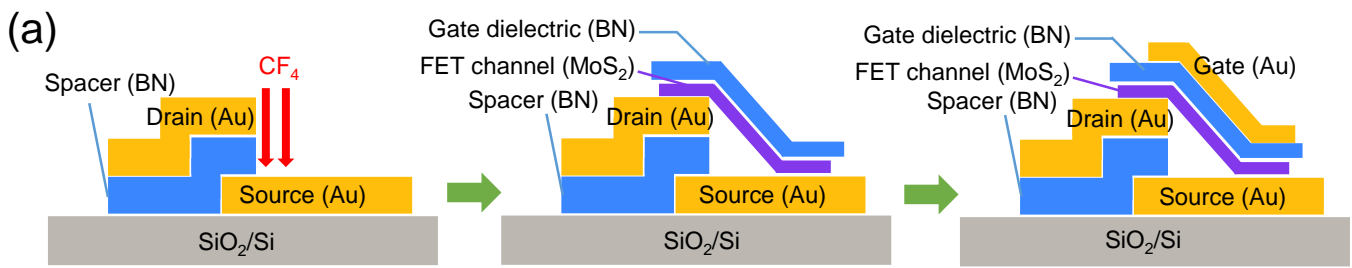
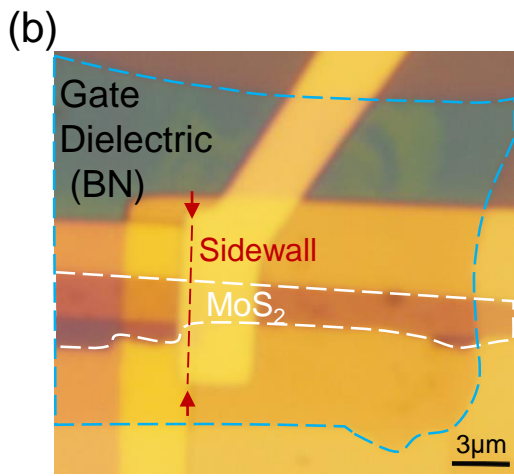
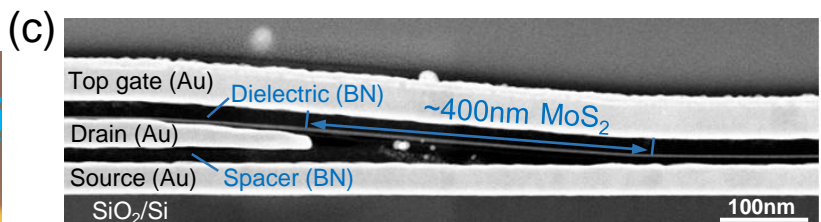
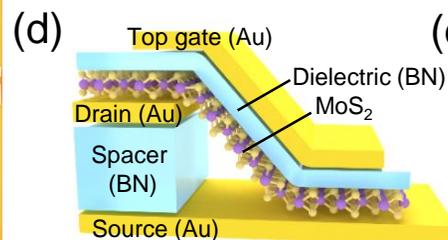
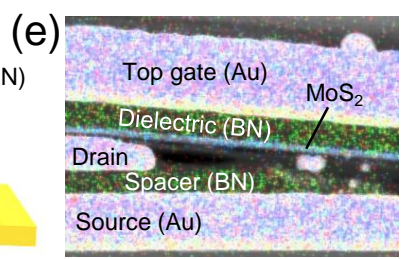
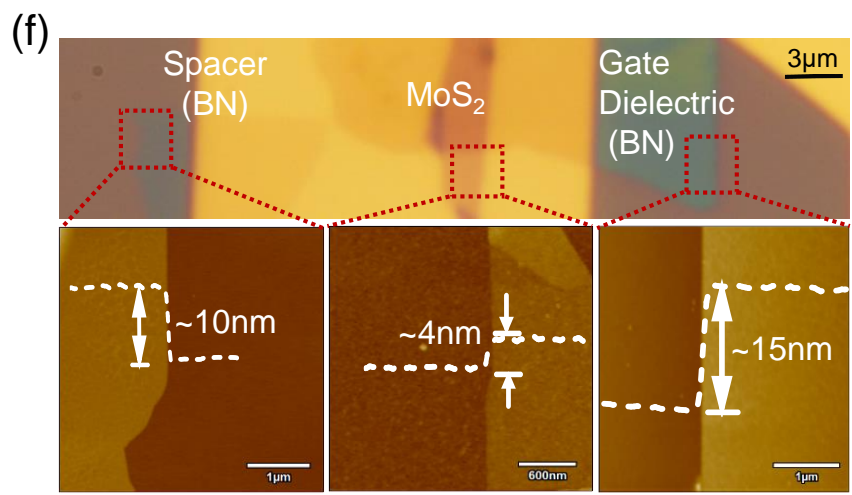
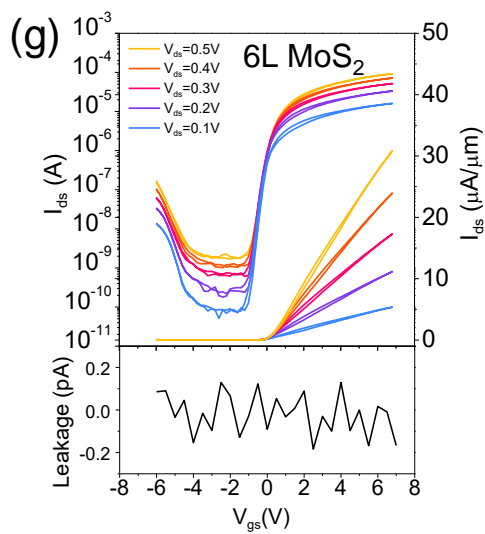

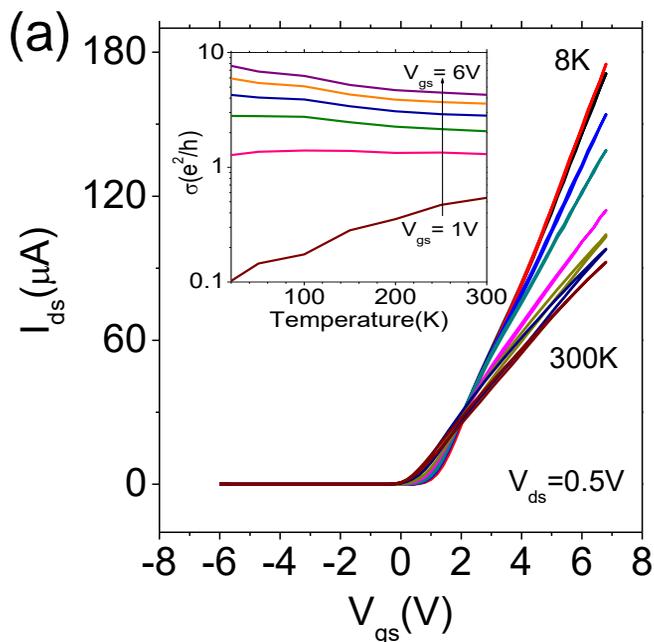
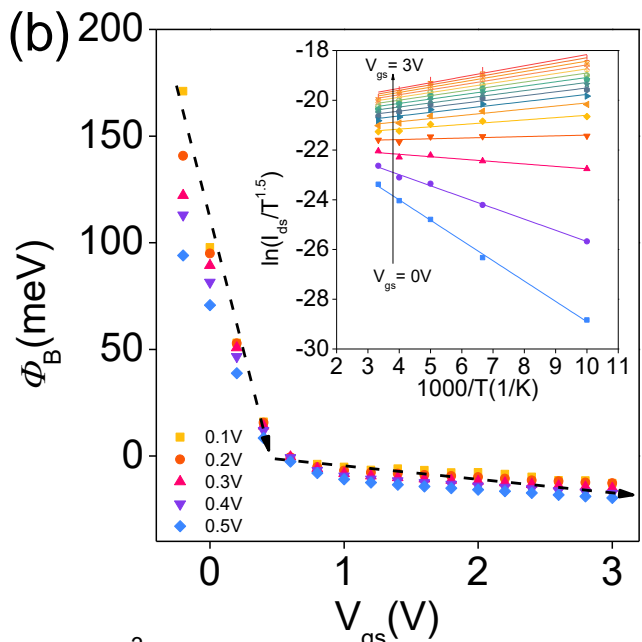
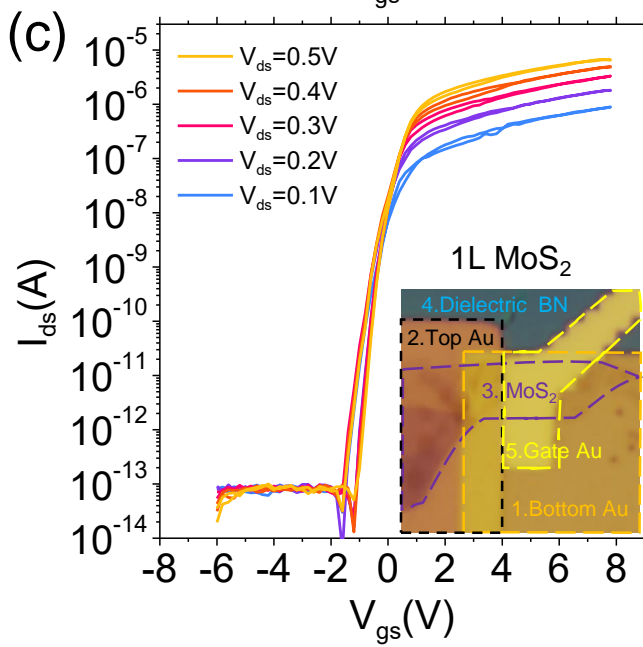
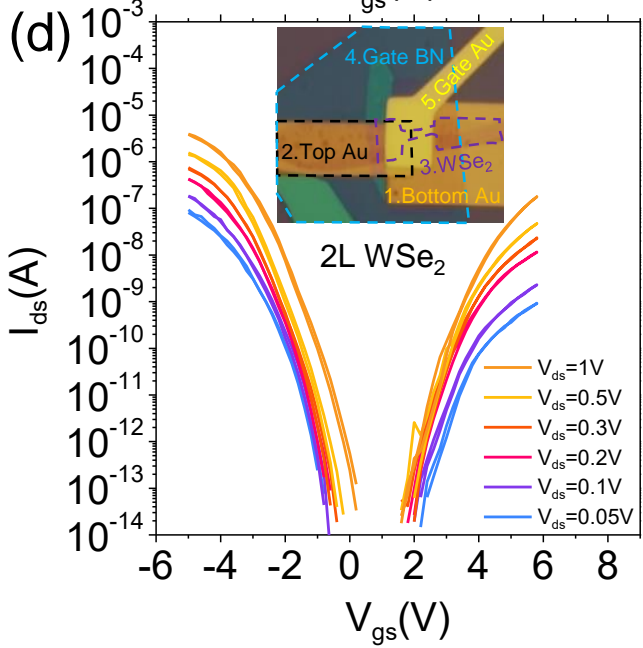

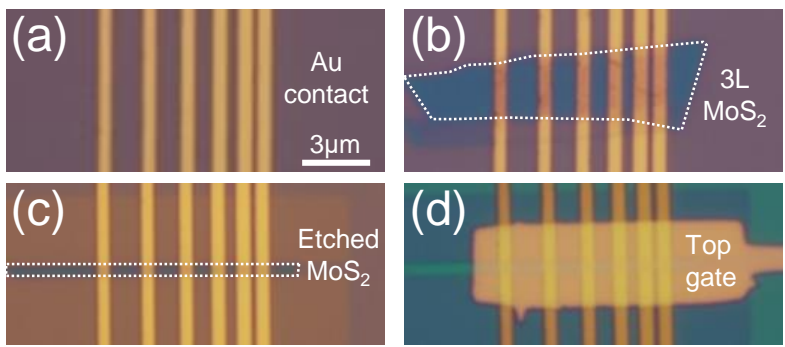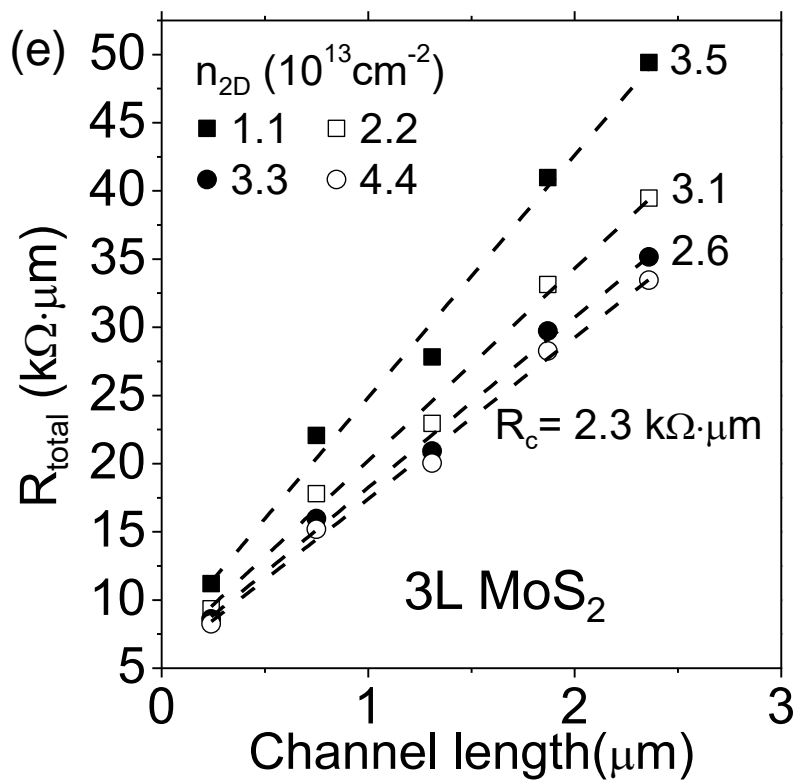

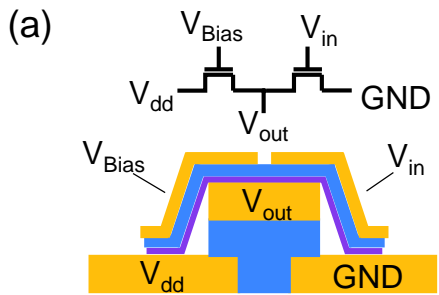
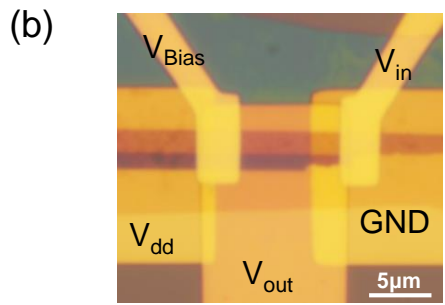
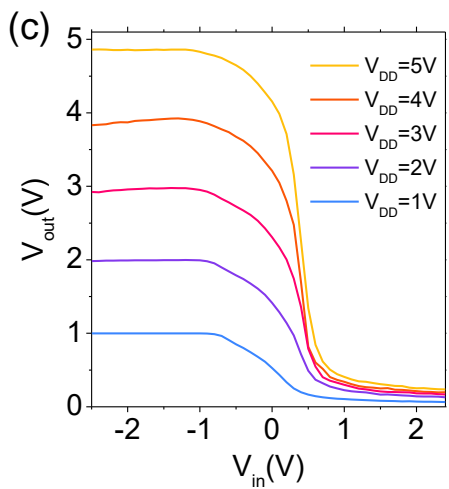
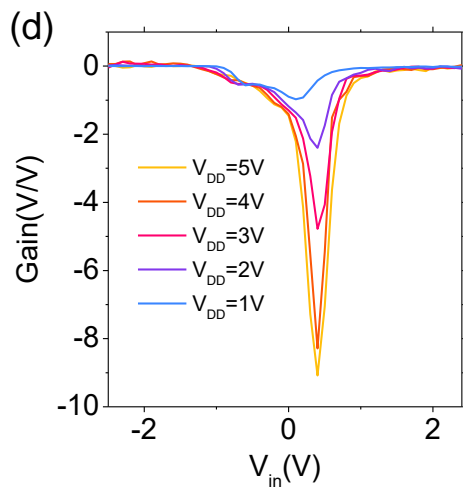

# A nanogapped hysteresis-free field-effect transistor

Jiachen Tang, Luhao Liu, Yinjiang Shao, Xinran Wang, Yi Shi and Songlin Li[a]

*National Laboratory of Solid-State Microstructures, Collaborative Innovation Center of Advanced Microstructures, and School of Electronic Science and Engineering, Nanjing University, Nanjing, Jiangsu 210023, China*

[a] *Author to whom correspondence should be addressed: sli@nju.edu.cn*

**Device fabrication**

The processing flow of the nanogapped FETs, which comprises 6 main steps, and corresponding optical images at each step are depicted in Fig. S1. The main steps are:

(a) Bilayer electron resists comprising methyl methacrylate (MMA, EL9) and poly methyl methacrylate (PMMA, A4) were spin coated onto the $SiO_2$/Si substrates. After patterning source electrode regions via electron beam lithography, the samples were then developed for 40 s and carried into a thermal evaporation chamber for metallization. Afterwards, Ni/Au (2 nm/30 nm) source electrodes were defined using a combination of thermal evaporation deposition and standard lift-off.

(b) The spacer BN flakes were mechanically exfoliated from bulk crystals with Scotch tape onto the polydimethylsiloxane (PDMS) substrates. Then, a suitable BN flake was selected and transferred onto predefined source electrodes.

(c) Ni/Au (2 nm/30 nm) drain electrodes were deposited onto the top of the stack of spacer BN and source electrode through the standard EBL and metallization as described in step (a). A small overlap between the top and bottom Au electrodes was intentionally left to fabricate the sidewall in the next step.

(d) Adopting the top drain electrode as a self-aligned mask, the exposed part of the spacer BN was etched away with inductively coupled plasma (ICP), using $CF_4$ as the etching media. Then, a sharp sidewall is formed, in which the top and bottom Au electrodes are electronically separated by the middle BN spacer.

(e) Van der Waals stacking of the BN-supported MoS$_2$ channel. Individual flakes of BN and 2D semiconductor channels (mainly metal dichalcogenides, e.g., MoS$_2$ and WSe$_2$) were initially mechanically exfoliated onto SiO$_2$/Si substrates. Then, a BN flake was selected as top encapsulator and picked up with a thick polydimethylsiloxane (PDMS) slab coated with polypropylene carbonate (PPC) polymer at 40 °C. Then, the BN/PPC/PDMS trilayer slab was used to further pick up the 2D channel and to transfer the 2D channel/BN bilayer to the predefined Au/BN/Au sidewall in step (d). During transfer, the whole sample was heated to 120 °C to make PPC melt and detach from the PDMS slab to release the 2D channel/BN bilayer. Next, the as-stacked sample was cleaned with acetone and was further annealed at 350 °C for 120 min in Ar/H$_2$ atmosphere to remove the bubbles between the layers.

(f) A Ni/Au (2 nm/60 nm) gate electrode was deposited onto the top of sidewall through the standard EBL and metallization as described in step (a).

**Y-function method**

The Y-function method is based on the straightforward analysis of the drain current ($I_{ds}$) in the strong inversion regime. In such a condition, the height and width of the Schottky barrier are basically fixed without considerable changes. Thus, contact resistance may be considered constant. The basic idea of the Y-function method is to extract contact resistance ($R_C$) through a derived Y function, based on a smart transformation of some relevant quantities.

Considering the mobility decays with increasing gate voltage which induced by remote phonon scattering and surface roughness, the $I_{ds}$-$V_{gs}$ equation can be expressed as

$$I_{ds} = \left(\frac{\mu_0}{1+\theta(V_{gs}-V_t)}\right)\frac{WC_{ox}V_{ds}}{L}(V_{gs}-V_t), \tag{S1}$$

where $\theta$ is the first-order mobility attenuation coefficient, $C_{ox}$ is the gate capacitance, $\mu_0$ is the intrinsic mobility, and $L$ and $W$ are device length and width. And according to voltage-division of contact resistance, the $I_{ds}$-$V_{gs}$ equation can also be expressed as

$$I_{ds} = \left(\frac{\mu_0}{\frac{R_{ch}+R_C}{R_{ch}}}\right)\frac{WC_{ox}V_{ds}}{L}(V_{gs}-V_t), \tag{S2}$$

where $R_{ch}$ denotes the channel resistance. By comparing formulae (S1) and (S2), $\theta$ can be written as

$$\theta = R_C \frac{W C_{ox} \mu_0}{L}, \tag{S3}$$

and the transconductance $G_m = dI_{ds}/dV_{gs}$ can be expressed as

$$G_m = \frac{W C_{ox} V_{ds} \mu_0}{L} \frac{1}{[1+\theta(V_{gs}-V_t)]^2}. \tag{S4}$$

Then $Y$-function and $G_m^{-0.5}$ can be written as

$$Y = \frac{I_{ds}}{G_m^{0.5}} = (\frac{W C_{ox} V_{ds} \mu_0}{L})^{0.5}(V_{gs} - V_t), \tag{S5}$$

$$G_m^{-0.5} = (\frac{W C_{ox} V_{ds} \mu_0}{L})^{-0.5}[1 + \theta(V_{gs} - V_t)]. \tag{S6}$$

The data points of the calculated $Y - V_{gs}$ curves and $G_m^{-0.5} - V_{gs}$ curves were linearly fitted to obtain the slope $K_1$ and $K_2$ as

$$K_1 = (\frac{W C_{ox} V_{ds} \mu_0}{L})^{0.5}, \tag{S7}$$

$$K_2 = \theta(\frac{W C_{ox} V_{ds} \mu_0}{L})^{-0.5}. \tag{S8}$$

By combining formulae (S3), (S7) and (S8) and considering the presence of two contact areas, one can derive the realistic $R_C = 0.5 V_{ds} W K_2 / K_1$.

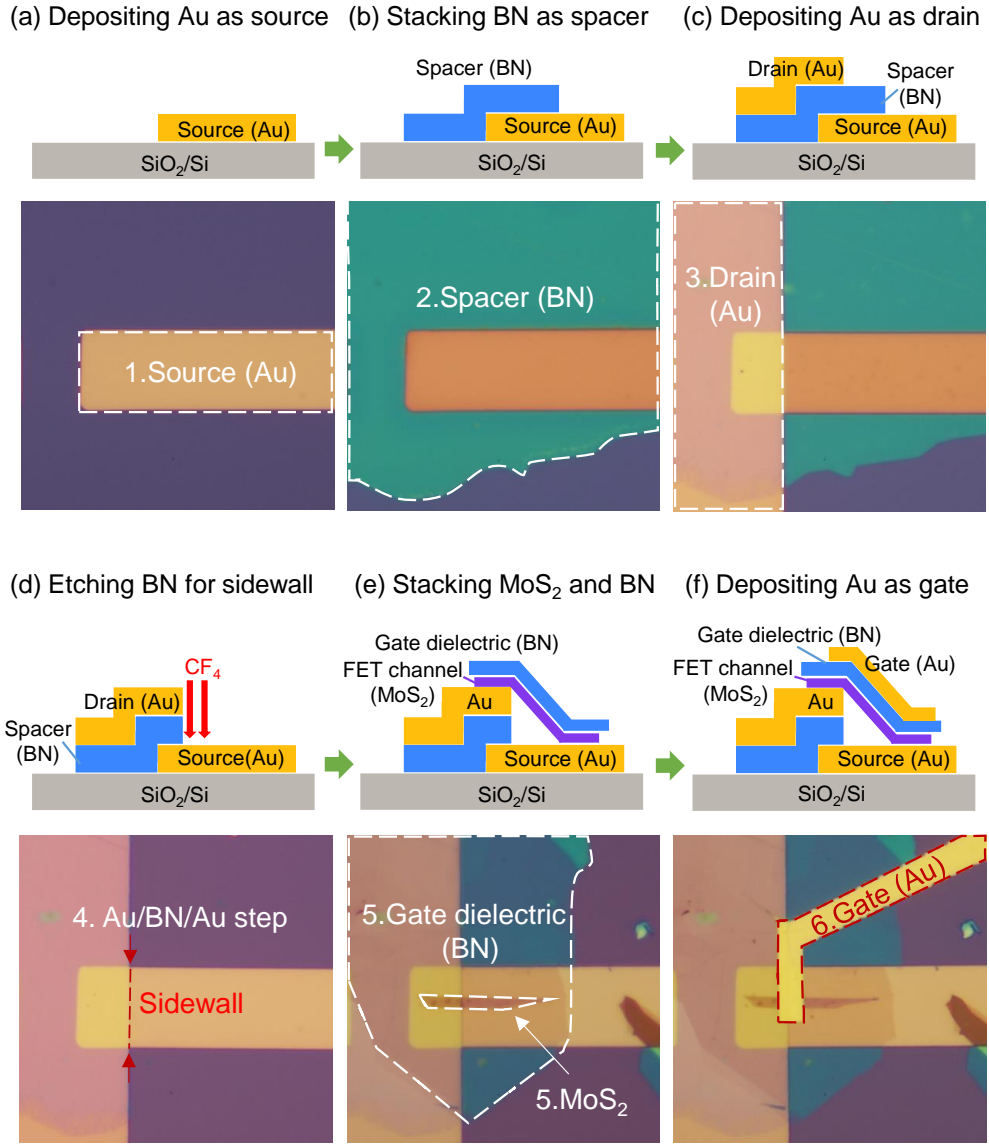

**FIG. S1.** Schematic diagrams and corresponding optical images taken each fabrication step to elucidate the device structure. (a) Depositing the bottom Au source electrode. (b) Stacking the BN as spacer onto the bottom Au electrode. (c) Depositing the top Au drain electrode. (d) Self-aligned etching of the exposed area of the BN spacer to form the sharp Au/BN/Au sidewall. (e) Van der Waals stacking the BN-supported $MoS_2$ channel onto the Au/BN/Au sidewall. (f) Depositing the Au top gate to finalize the fabrication.

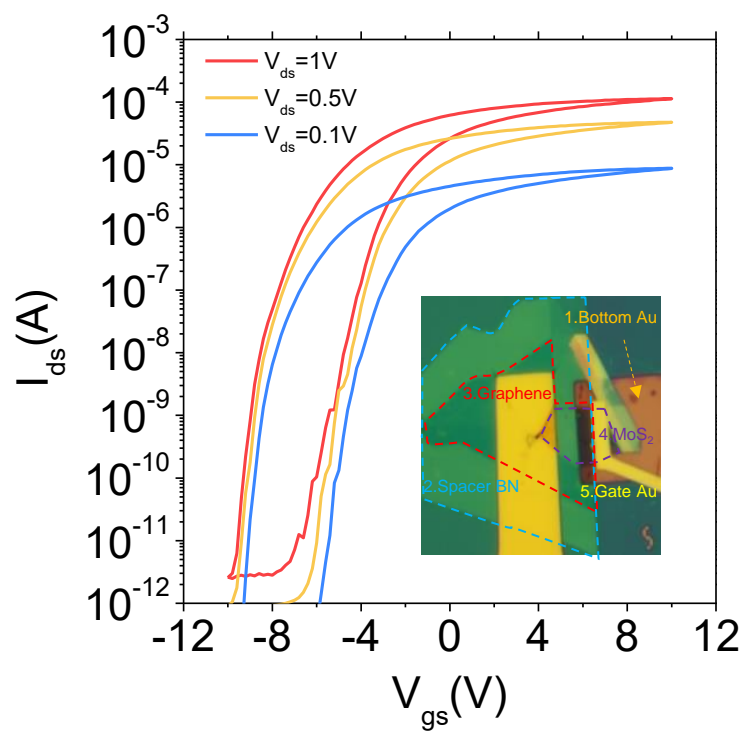

**FIG. S2.** Transfer characteristics for a semi-suspended FET with 20nm $Al_2O_3$ as the gate dielectric, where noticeable scanning hysteresis arises. Inset: optical image for the FET.

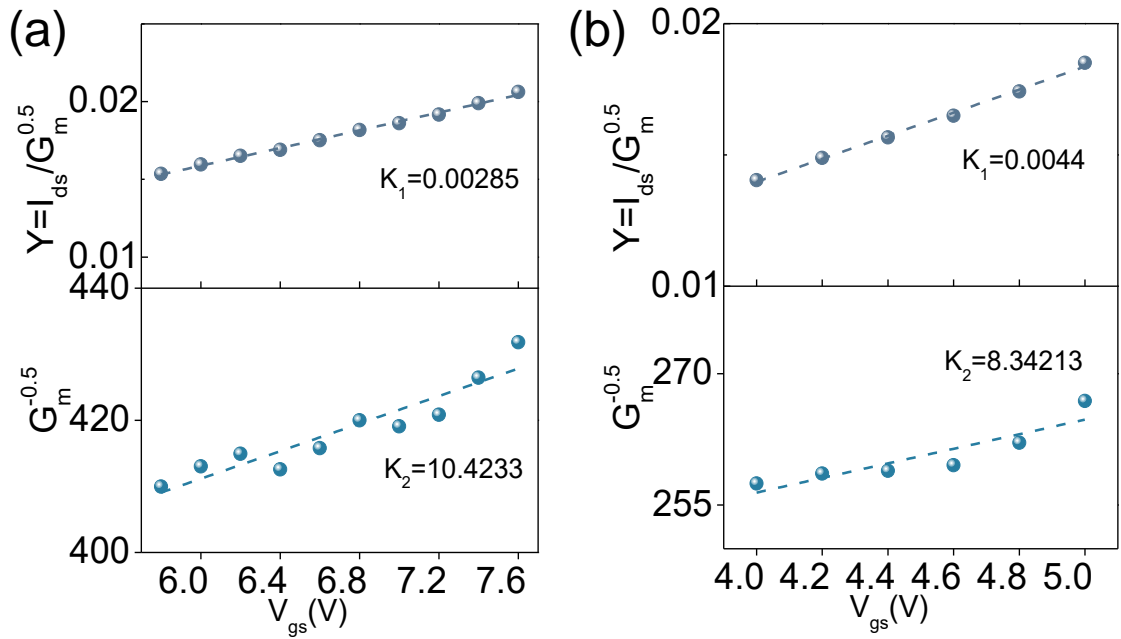

**FIG. S3.** $R_C$ estimation using the Y-function method for 8L and 6L MoS$_2$ semi-suspended FETs. Extracting the slopes from the $V_{gs}$ dependent Y function and the square root of transconductance for FETs with (a) 8L and (b) 6L MoS$_2$ channels.

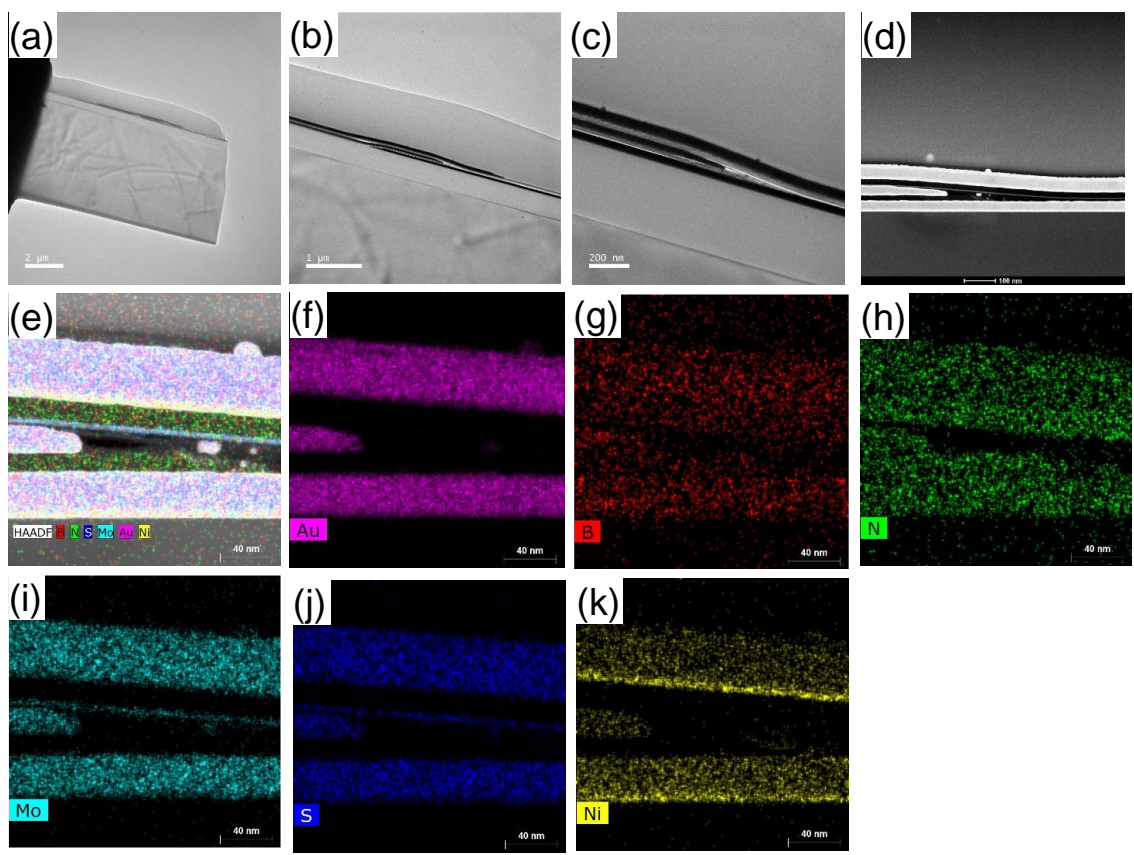

**FIG. S4.** (a)−(d) Cross-sectional TEM images for a typical nanogapped FET taken at different magnification ratios. (e)−(k) Elemental mappings around the MoS$_2$ channel area. The names of the main elements of the FETs are given in the bottom left corner of each panel.